\documentclass[preprint,showpacs,preprintnumbers,amsmath,amssymb]{revtex4}

\usepackage[dvips]{graphicx}

\newcommand{\etal}{\textit{et~al.}}

\begin{document}

\title{Dephasing of Mollow Triplet Sideband Emission of a Resonantly Driven Quantum
Dot in a Microcavity}

\author{S. M. Ulrich$^1$\footnote{Corresponding author:
s.ulrich@ihfg.uni-stuttgart.de}, S. Ates$^1$, S. Reitzenstein$^2$,
A. L\"offler$^2$, A. Forchel$^2$, and P. Michler$^1$}

\affiliation{$^{1}$Institut f\"{u}r Halbleiteroptik und
Funktionelle Grenzfl\"a{c}hen $\&$ Research Center SCoPE,
Universit\"at Stuttgart, D-70569, Stuttgart, Germany}

\affiliation{$^{2}$Technische Physik, Wilhelm Conrad R\"ontgen
Research Center for Complex Material Systems, Universit\"at
W\"urzburg, Am Hubland, D-97074 W\"urzburg, Germany}

\begin{abstract}
Detailed properties of resonance fluorescence from a single
quantum dot in a micropillar cavity are investigated, with
particular focus on emission coherence in dependence on optical
driving field power and detuning. Power-dependent series over a
wide range could trace characteristic Mollow triplet spectra with
large Rabi splittings of $|\Omega| \leq 15$~GHz. In particular,
the effect of dephasing in terms of systematic spectral broadening
$\propto \Omega^2$ of the Mollow sidebands is observed as a strong
fingerprint of \textit{excitation-induced dephasing}. Our results
are in excellent agreement with predictions of a recently
presented model on phonon-dressed QD Mollow triplet emission in
the cavity-QED regime.
\end{abstract}

\pacs{78.67.Hc, 42.50.Dv, 78.55.-m}

\maketitle
Exploiting the quantum properties of the light which is emitted
from semiconductor quantum dots (QDs) has the potential of
enabling various new applications in the field of photonics and
quantum information technology~\cite{Bouwmeester:2000}. Many of
these applications require single-\cite{Michler:2000,Yuan:2002}
and
entangled-photon~\cite{Gershoni:2006,Stevenson:2006,Hafenbrak:2007,Dousse:2010}
light sources. The implementation of single-photon based quantum
logic algorithms critically relies on photon
indistinguishability~\cite{Santori:2002} which is directly related
to the coherence properties of the emitted light. The coherence
time $T_2$ can be defined via the excited state's dephasing rate
$(T_2)^{-1} = (2 T_1)^{-1} + (T_2^\ast)^{-1}$, with $T_1$ as the
radiative emitter lifetime and $T_2^\ast$ as the pure dephasing
time. True resonant s-shell excitation appears beneficial to avoid
pure dephasing, and is therefore highly anticipated to approach
the ideal Fourier transform limit~\cite{Kiraz:2004,Ates:2009}
given by $T_2 / (2T_1) = 1$. The observation of the Mollow triplet
is the characteristic hallmark of resonance fluorescence from a
strongly driven and dressed two-level system. Resonance
fluorescence emission from a single QD has been first reported by
Muller et al.~\cite{Muller:2007}. Recently, they also directly
observed Mollow triplets~\cite{Mollow:1969} with Rabi splittings
up to $\Omega \approx 4.4$~GHz ($18\,\mu$eV), where a distinct
pump-power independent dephasing rate of $\Gamma_2^\ast =
(T_2^\ast)^{-1} \approx 3.1$~GHz has been found~\cite{Flagg:2009}.
In another remarkable work by Vamivakas
\etal~\cite{Vamivakas:2009} the resonance fluorescence and
spin-resolved characteristics of a trionic QD state have been
investigated. The Mollow triplet sideband emission showed a high
degree of coherence with $\Gamma_2^\ast \approx 18$~MHz. Important
to note, all these investigations have been performed under
moderate pumping levels ($\Omega < 6$~GHz) and investigations
under stronger s-shell excitation have not been reported so far.

In this letter, we report on detailed investigations of resonance
fluorescence emission from single QD neutral excitonic $X^0$
recombination in a high-$Q$ micropillar cavity. The influence of
(a) \textit{resonant pump power} and (b) \textit{frequency
detuning of the driving laser} (-3~GHz~$< \Delta <$~4~GHz) on the
characteristic Mollow sideband emission is investigated, with
particular focus on the coherence properties of the emission.

The sample structure under study is grown by molecular beam
epitaxy on GaAs substrate. The initial planar cavity consist of 30
(26) AlAs/GaAs distributed Bragg reflector period pairs as the
bottom (top) mirrors, respectively. Spacer of $2 \times 130$~nm
GaAs form a $\lambda$-cavity around a single centered layer of
(In,Ga)As QDs with a spatial density of $\sim 6 \cdot
10^{9}$~cm$^{-2}$. Ordered fields of different diameter ($1.5 -
4\,\mu$m) high-quality micropillars have been finally processed by
combined electron-beam lithography and plasma-induced reactive ion
etching~\cite{Reitzenstein:2007}.

Our investigations are performed on a confocal low-temperature ($T
\geq 4$~K) micro-photoluminescence ($\mu$-PL)
setup~\cite{Ates:2009}, using a special orthogonal symmetry
between lateral optical laser excitation of individual micropillar
structures close to the cleaved $[110]$ sample edge and vertical
QD emission detection along the pillar symmetry axis. A
narrow-band (FWHM~$\approx 500$~kHz) continuous-wave (cw)
Ti:Sapphire ring laser or by a mode-locked Ti:Sa pulse laser
($\sim 2$~ps pulses at $f_{rep} = 76.2$~MHz) for time-resolved
$\mu$-PL (TCPC: time-correlated photon counting) are used for
excitation. In addition to $\mu$-PL detection by a
spectrometer/CCD system ($\Delta E_{res} \sim 35\,\mu$eV
resolution), a scanning Fabry-P\'{e}rot
interferometer~\cite{Ates:2009} (Finesse~$F \sim 150$; FSR~=~$15\,
\text{GHz} \sim 62.035\,\mu$eV free spectral range) provides
high-resolution PL (HRPL) with $\Delta E_{res}^{\text{HRPL}} <
1\,\mu$eV.


The presented measurements have been performed on a 1.75~$\mu$m
diameter micropillar. As shown in Fig.~1(a), this microcavity
reveals single-QD neutral exciton $X^0$ s-shell emission at $\sim
1.3571$~eV close to the FM ($\sim 1.3568$~eV) at $T = 10$~K under
selective \textit{quasi-resonant} QD p-shell excitation $\sim
22$~meV above the ground state. At this QD-FM detuning $\Delta E =
+280\,\mu$eV, the spontaneous radiative $X^0$ decay time has been
measured as $\tau_{dec} = 820\,\pm\,40$~ps by TCPC under pulsed
p-shell excitation (see inset of Fig.~1(a)). Due to very fast
carrier relaxation between the p- and s-shell in those QDs (typ.
on a few ps-scale), the pure radiative lifetime $T_1$ is expected
to obey $T_1 \approx \tau_{dec}$. From the ratio of FM emission
energy and linewidth $\delta E$ (FWHM), we derive a high quality
factor of $Q \approx E/\delta E = 13500\,\pm\,500$ for this
micropillar. Worth to note, the prominent effect of fundamental
mode emission despite its distinct spectral \textit{detuning} from
the dominantly 'feeding' single QD is due to \textit{non-resonant
QD-mode coupling} in such solid-state emitter-cavity systems
\cite{Hennessy:2007,Press:2007,Kaniber:2008,AtesUlrich:2009}.


As was first theoretically described by
B.~R.~Mollow~\cite{Mollow:1969}, a strong and resonant driving
light field (with detuning $\Delta = 0$) 'dresses' the electronic
states of a two-level system into a four-level system (Fig.~2(a)).
Optical emission of such a system is described by characteristic
multi-Lorentzian \textit{Mollow triplet} spectra, composed of a
central peak at the bare emitter frequency $\nu_0$, symmetrically
decorated by two satellites at $\nu_0 \pm \Omega$ (see Fig.~2(b)).
$\Omega = \mu E_0/h$ represents the bare Rabi frequency (in
\textit{Hz}) with transition dipole moment $\mu$ and the local
field strength $E_0$ at the emitter. For $\Delta = 0$ and
sufficient excitation, the energetic sideband splitting with
respect to $\nu_0$ is given by $\sim |\Omega|$, i.e., obeys a
proportionality to the square root of excitation
power~$P_0^{1/2}$.

Results of systematic HRPL investigations on single QD exciton
s-shell emission under strictly resonant ($\Delta = 0$),
power-dependent cw excitation are depicted in Fig.~2(b). For
enhanced clarity, the spectra are vertically shifted. Spectral
detuning of emission relative to the bare $X^0$ transition is
denoted as $\delta$ in units of frequency (GHz) and energy
($\mu$eV). As a consequence of the Fabry-P\'{e}rot technique used
in HRPL, all spectral emission features appear periodically with
an offset equal to the interferometer FSR. With increasing power,
the evolution of two symmetrically spaced sidebands can be clearly
observed around the central line which is composed of QD resonance
fluorescence and residual scattered laser light. By variation of
the pump power over more than two orders of magnitude, a
systematic increase of Rabi sideband splittings up to $\Omega
\approx \pm \, 15$~GHz ($\pm\,62\mu$eV) is traced, limited only by
the FSR of our interferometer, as here the sidebands start to
overlap with adjacent FPI transmission orders. Extracted side peak
splittings $|\Omega|$ from Fig.~2(b) are plotted in Fig.~2(c) as a
function of the square root of the driving laser power ($\sim
P_0^{1/2}$). The theoretically expected proportionality is clearly
confirmed, as demonstrated by the applied linear fit (solid line).
We like to emphasize that in our studies the investigated
excitation power range and consequently the regime to observe Rabi
splitting values of $|\Omega|$ is significantly larger than in
previously reported investigations
\cite{Flagg:2009,Vamivakas:2009} using direct detection of
'dressed' state fluorescence from single QDs.

From a more detailed inspection of the Mollow triplet series in
Fig.~2(b), we particularly observe also a systematic and distinct
spectral broadening of the Rabi sidebands with increasing pump
strength. For preliminary analysis, the spectral line widths
$\Delta \nu$ and their gradual broadening have been deduced from
Lorentzian least-square fits to the HRPL data (not shown).
Figure~2(d) (black trace) depicts corresponding FWHM values (GHz)
of the Rabi sidebands, which reveal an overall line width increase
by factor $\sim 1.8$ over the observed power range. In particular,
a clear linear dependence $\Delta \nu \propto \Omega^2$ on the
squared Rabi frequency is traced, which represents a strong
indication of \textit{excitation-induced dephasing} (EID) as an
important additional effect accompanying the 'dressed' character
of resonant QD emission. As is shown below, our experimental
findings are in full quantitative consistence with predictions of
a recently presented model on phonon-dressed QD Mollow triplet
emission in the cavity-QED regime~\cite{RoyArXiv:2011}.

Prior to detailed data analysis, we note here that other studies
on \textit{pulsed resonant} single QD excitation have interpreted
the effect of EID in terms of (time-domain) Rabi rotation damping
to originate from coherent energy exchange between the emitter and
a resonant LA-phonon bath in the barrier matrix
\cite{Ramsay:2010,RamsayArXiv:2010}. In addition to pulsed
broadband emission, their model addressed InGaAs QDs in
$n-i$-Schottky diode structures \textit{without} a surrounding
micro cavity -- in clear contrast to our experimental conditions.
Alternatively, carrier scattering between a resonantly
\textit{pulsed} single QD in a photodiode structure (without
cavity coupling) and off-resonant wetting layer and multi-exciton
states were theoretically discussed as the origin of EID with
respect to decoherence in Rabi
oscillations~\cite{Villas-Boas:2005}. Moreover, for resonance
fluorescence under pure \textit{continuous wave} (cw) conditions,
photon statistics and emission dephasing via phonon-bath coupling
of a single bare QD without cavity coupling have also been
analyzed theoretically~\cite{Nazir:2008}. In qualitative
consistence of all previous
studies~\cite{Villas-Boas:2005,Nazir:2008,Ramsay:2010,RamsayArXiv:2010},
EID rates proportional to the squared Rabi frequency $\Omega^2$
were concluded. Nevertheless, the particular regime of cavity-QED
providing distinct emitter-mode coupling between a single
'dressed' QD and a high-Q 3D micro cavity has not been addressed
by either previous study, thus hindering a direct numerical
interpretation of our data in Fig.~2. In contrast, we will compare
our results to a new recently presented theory by Roy and
Hughes~\cite{RoyArXiv:2011} on cw-excited single QD resonance
fluorescence in a high-Q cavity, which explicitly considers the
combined effects of electron-acoustic phonon bath coupling and
electron-photon coupling to model a full \textit{cavity-QED
system} on the basis of polaronic dynamics.

To quantitatively interpret the observed spectral sideband
broadening (Fig.~2(b)), our Mollow triplet spectra were modeled on
the theoretical basis of Ref.~\onlinecite{Flagg:2009}. Although
the authors report no EID effect from their experiments, the
influence of pure dephasing was already accounted for by
independent rates of radiative decay $(2T_1)^{-1}$ and decoherence
$\Gamma = T_2^{-1}$ (Ref.~\onlinecite{Flagg:2009}, Eqn.~3). In our
analysis, the total dephasing rate was substituted as $\Gamma =
(2T_1)^{-1} + K \cdot \Omega^2$ to compute the side peak spectra
in dependence on \textit{radiative dephasing} and power-dependent
\textit{pure dephasing} (EID) with emission broadening $\Delta \nu
\propto \Omega^2$. For various fixed values of $K$ (GHz$^{-1}$),
the whole set of Mollow triplet side peak spectra was repeatedly
calculated~\cite{footnote}. High consistence over the full power
series could be achieved for a dephasing parameter $K = 0.005 \pm
0.001$~GHz$^{-1}$ (at $T = 10$~K), well reproducing all side peak
spectra (Fig.~2(b), bold red lines) and their overall FWHM
broadening, i.e. the according dephasing rate $\Gamma$ extracted
in Fig.~2(d). Worth emphasizing, apart from radiative dephasing no
extra constant pure dephasing needed to be included, which
anticipates close to ideal Fourier transform-limited resonance
fluorescence $T_2/(2 T_1) \approx 1$ in the limit $\Omega^2
\rightarrow 0$, in full agreement with our previous studies on
single-QD resonance emission \cite{Ates:2009}.

To numerically compare the observed Mollow side peak broadening
with respect to predictions of the QED model and the coupling
regime considered in Ref.~\onlinecite{RoyArXiv:2011}, we deduced
the emitter-cavity coupling strength $g$ for the studied QD-cavity
system from $g \approx \sqrt{\frac{F_P \, \cdot \, \kappa_{cav}}{4
\, \tau_X}}$. For a measured Purcell emission enhancement factor
of $F_P \approx 13$, a FM cavity photon loss rate of $\kappa =
\frac{2 \pi c_0}{Q \cdot \lambda_X} \approx 104\,\mu$eV (i.e.
$\sim 25.2$~GHz) and the measured excitonic emission life time of
$\tau_X = 820 \pm 40$~ps at $\Delta E = +280\,\mu$eV QD-FM
detuning (Fig.~1(a)), we obtain a value of $g \approx 16.2 \pm
0.5$~$\mu$eV (or $3.9 \pm 0.1$~GHz), conform with the regime
described by Roy and Hughes~\cite{RoyArXiv:2011}. Their
calculations, explicitly considering pure cw-resonant s-shell
excitation under zero laser detuning ($\delta = 0$), clearly
predict the effect of excitation-induced dephasing (EID) due to
combined emitter phonon-bath coupling and cavity-photon bath
interaction especially in those cavity QED systems. In particular,
distinct spectral Mollow triplet side band broadening $\Delta \nu
\propto \Omega^2$ is expected for small and moderate
emitter-cavity couplings in the range $15 \,\mu\text{eV} \leq g
\leq 50\,\mu\text{eV}$. Considering parameters very similar to
those valid for our experiments, Mollow side peak bandwidth
enhancement with $K \approx 0.005 - 0.007$~GHz$^{-1}$ (equivalent
to $K \approx 0.001 - 0.002$~$\mu$eV$^{-1}$) is indeed
anticipated, in high quantitative conformity with our experimental
results discussed above (Fig.~2(d)).

Furthermore, as a direct consequence of phonon-mediated
non-resonant cavity feeding from the detuned QD into the detuned
FM ($\Delta E = +280\,\mu$eV here), an asymmetry between the
Mollow triplet side band emission intensities is
expected~\cite{RoyArXiv:2011} due to their slightly different
spectral detuning to the micro cavity mode. Clear indications of
this effect could be consistently traced from the detailed
theoretical fit of our HRPL series in Fig.~2(a) (bold lines),
reflecting $\sim 15\%$ higher intensity from the low frequency
Mollow triplet side bands at $\nu_0 - \Omega$ (i.e. spectrally
closer to the FM) with respect to their counterparts at $\nu_0 +
\Omega$.

The second part of our studies focused on single-QD Mollow triplet
spectra in explicit dependence on laser frequency detuning.
According to theory~\cite{Vamivakas:2009}, non-zero spectral
detuning $\Delta$ between the driving field and the bare emitter
resonance $\nu_0$ characteristically modifies the 'dressed'
emission. Besides the center transition at $\nu_0 + \Delta$, the
two sideband frequencies become $\nu_0 + \Delta \pm
\Omega^{\prime}$, where $\Omega^{\prime}$ denotes the
\textit{generalized Rabi frequency} $\Omega^{\prime} =
\sqrt{\Omega^2 + \Delta^2}$ with $\Omega$ as the (zero detuning)
'bare' Rabi frequency at a given excitation power.


Figure~3(a) depicts HRPL spectra at a fixed laser power of $P_0
\approx 16\,\mu$W, taken under systematic detuning of the
excitation frequency from the QD s-shell resonance over a large
range of $-3\,\text{GHz} \leq \Delta \leq \,+4\,\text{GHz}$
(indicated by arrow markers). The center peak composed of QD
resonance fluorescence and intense laser stray light has been
removed from the spectra for clarity. Under variation of $\Delta$,
a gradual shift of the whole Mollow triplet signature and a
significant increase of sideband splitting with increasing
detuning are clearly observable. Figure~3(b) extracts the peak
positions of each emission component in Fig.~3(a) as a function of
spectral laser detuning $\Delta$. As a reference, the horizontal
dashed line denotes the symmetry point $\Omega' = \Omega$ of
zero-detuning from the QD s-shell. Explicit fits of the sideband
positions according to $\nu(\Delta) = \nu_0 + \Delta \pm
\sqrt{\Omega^2 + \Delta^2}$ (Fig.~3(b), bold red lines) reveal
high consistence with theory. For a more convenient analysis of
the detuning series, values of the \textit{full sideband
splittings} as a function of $\Delta$ are plotted in Fig.~3(c),
revealing a minimum for $\Delta \rightarrow 0$. From a
least-square fit $\propto 2\,\sqrt{\Omega^2 + \Delta^2}$ to the
data (bold line) at this fixed power level, we derive a bare Rabi
frequency of $\Omega = 3.99 \,\pm\,0.06$~GHz. Taking into account
this value of $\Omega$ together with idealized beam geometry and
spatial QD-field overlap for an estimation of the local field
strength $|E_{loc}|$, we deduce the magnitude of the electric
dipole moment of this particular QD as $\mu_{el} = h
\Omega/|E_{loc}| = 18 \pm 2$~Debye. This value is only somewhat
smaller than recently reported values for neutral $X^0$ states in
similar types of InAs/GaAs QDs~\cite{Vamivakas:2009}.

Another interesting phenomenon is traced from a detailed
evaluation of the sideband FWHM in the HRPL series of Fig.~3(a),
extracted in Fig.~3(d) (color-coded for the red (blue) detuned
Mollow triplet side bands). Under increasing laser detuning from
the QD s-shell resonance but fixed excitation power ($P_0 \approx
16\,\mu$W), we observe a distinct spectral \textit{narrowing} of
either side band with increasing $\Delta$. Worth noting, this
behaviour appears unexpected in terms of the cavity QED-based EID
model \cite{RoyArXiv:2011}, as 'dressed' emission with increased
side band splittings equivalent to the generalized Rabi frequency
$\sqrt{\Omega^2 + \Delta^2}$ should reflect also in increased
dephasing~\cite{RoyHughes:Note}, i.e. line broadening $\propto
\Omega'$ -- in contrast to our experiment. On the other hand,
first indications of abberations from the above denoted
theoretically expected emission detuning dependence are found by
detailed inspection of the extracted total side band splitting
with $\Delta$ in Fig.~3(c). With increasing detuning from
resonance we observe an increasing deviation between the measured
splittings and the theoretical model, which might tentatively be
interpreted to result from a systematically reducing bare Rabi
frequency independent of the increasing value of $\Delta$.
Nevertheless, a fundamental interpretation of this effect has to
be left for further ongoing in-depth experimental and theoretical
analysis.

In conclusion, resonance fluorescence emission proper\-ties of a
single InAs/GaAs QD in a high-quality micro\-pillar cavity have
been investigated in detail with particular focus on
\textit{excitation power} and/or \textit{excitation detuning}
relative to the s-shell ground state. Wide-range excitation power
series could trace characteristic Mollow triplet spectra with
large Rabi splittings of $\Omega \approx \pm 15$~GHz, accompanied
by the effect of systematic spectral sideband broadening $\propto
\Omega^2$ as a strong indication of \textit{excitation-induced
dephasing} (EID).

We gratefully acknowledge financial support by the DFG through the
research group 730 ''Positioning of Single Nanostructures --
Single Quantum Devices''.


\newpage
\textbf{Figures:}\\

\textbf{Fig.~1}

\begin{figure}[!ht]
\begin{centering}
\includegraphics[width=0.44\textwidth]{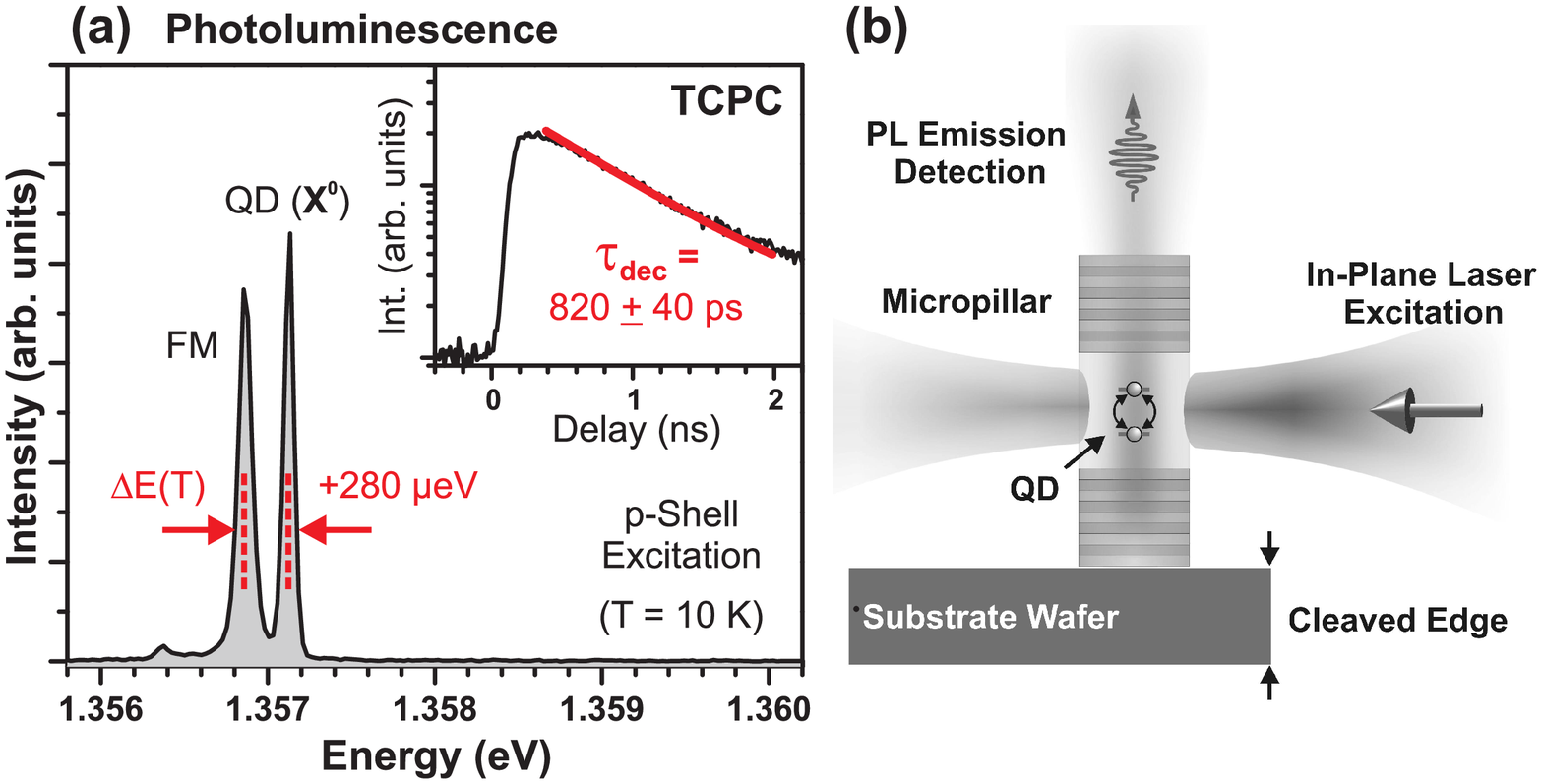}
\vspace{-0.3cm}\label{Fig1}
\end{centering}
\end{figure}

\vspace{1cm}
\textbf{Fig.~2}

\begin{figure}[!ht]
\begin{centering}
\includegraphics[width=0.37\textwidth]{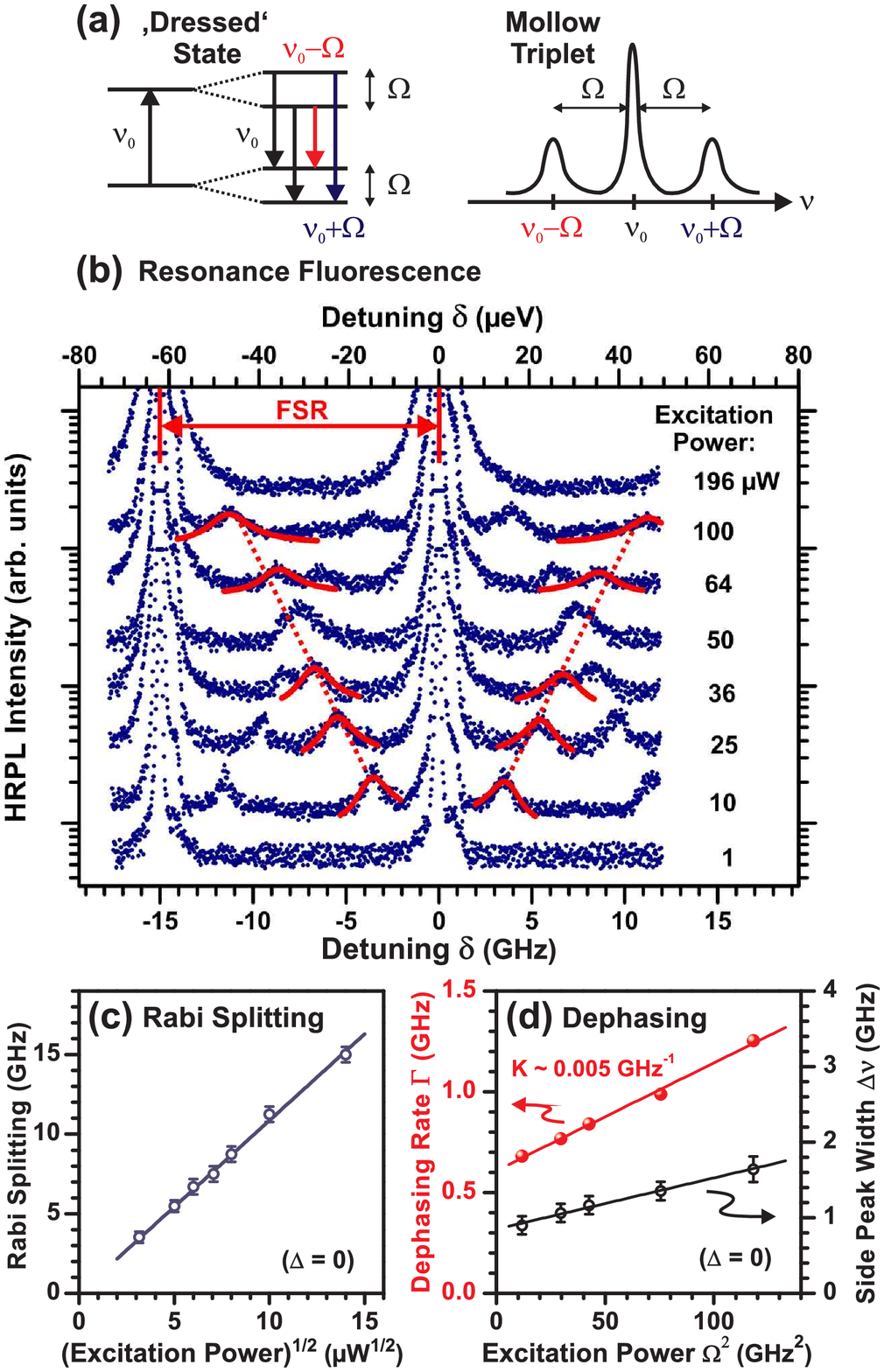}
\vspace{-0.3cm}\label{Fig2}
\end{centering}
\end{figure}

\newpage
\textbf{Fig.~3}

\begin{figure}[!ht]
\begin{centering}
\includegraphics[width=0.37\textwidth]{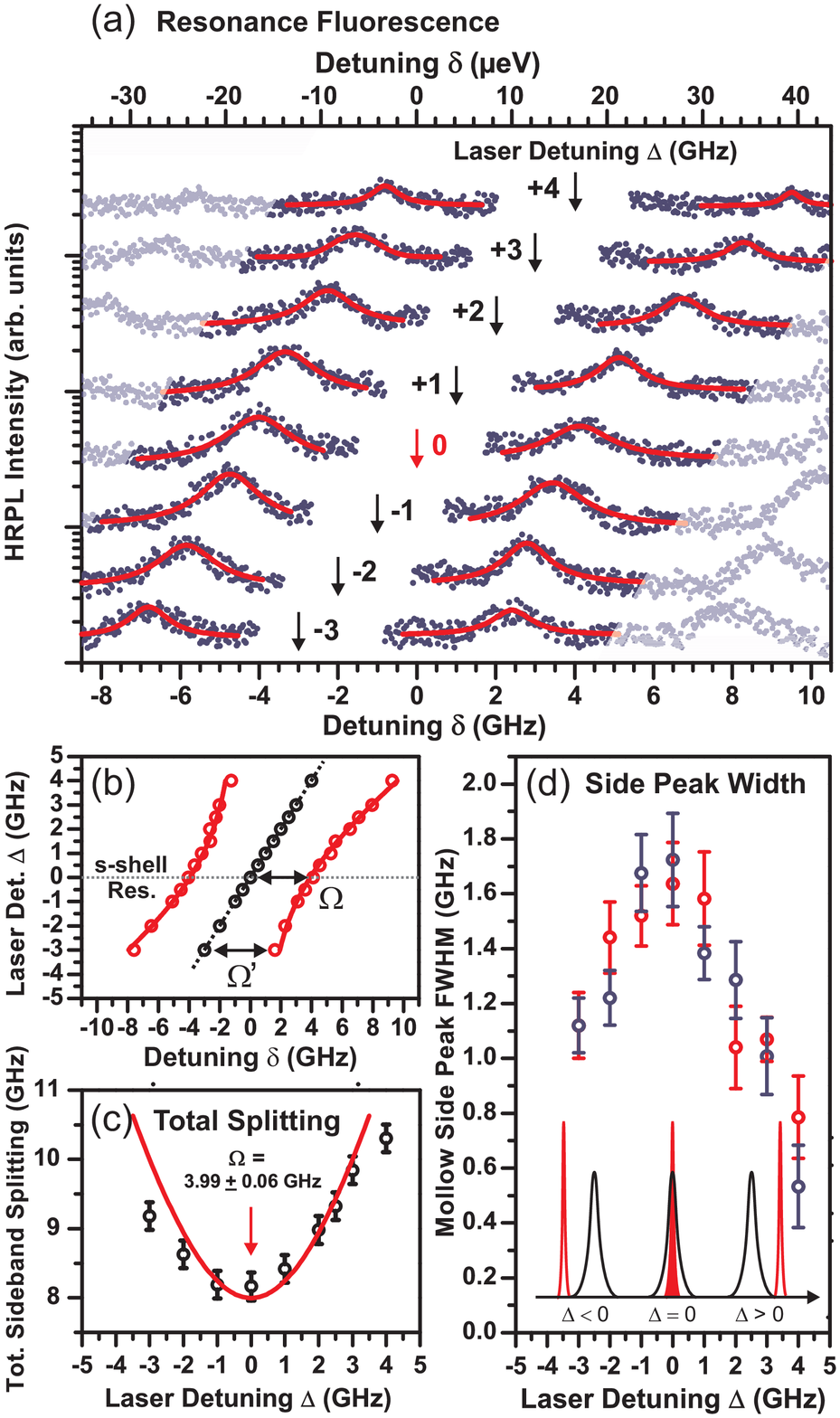}
\vspace{-0.3cm}\label{Fig3}
\end{centering}
\end{figure}

\newpage
\textbf{Figure Captions:}\\

\textbf{Fig.~1} (color online) (a) Single QD neutral exciton
($X^0$) emission spectrally close to the fundamental mode (FM) of
a $1.75\,\mu$m micropillar cavity, observed at quasi-resonant
p-shell excitation ($P_0 \approx 50 \mu$W; $\Delta E(T)$:
temperature-dependent QD-mode detuning). Inset: Time-resolved
$X^0$ emission signal. (b)
Schematic geometry of orthogonal excitation and detection.\\

\textbf{Fig.~2} (color online) (a) Energy scheme (left) and
characteristic Mollow triplet spectra (right) of a 'dressed'
two-level system. (b) Power-dependent high-resolution (HRPL)
spectra of $X^0$ emission at pure resonant ($\Delta = 0$)
continuous-wave excitation. (c) Mollow sideband splitting
extracted from plot~(b), revealing the expected linear dependence
on the square root of laser power $|\Omega| \sim
{P_0}^{1/2}$~(GHz). (d) Power-dependent Mollow sideband FWHM
$\Delta \nu$ and total dephasing rate $\Gamma \propto \Omega^2$
from theoretical analysis of data in (a),
indicative of excitation-induced dephasing (EID).\\

\textbf{Fig.~3} (color online) (a) HRPL of QD resonance
fluorescence under excitation detuning $\Delta$ (see arrows)
relative to the QD s-shell, taken at a fixed power of $P_0 =
16\,\mu$W. Red traces: Lorentzian FWHM fits. (b) Spectral
evolution of Mollow sidebands (red) with laser detuning $\Delta$
(black center trace), derived from (a). (c) Total sideband
splitting in plot~(b), together with a theory fit to evaluate the
bare Rabi frequency $\Omega$. (d) Sideband FWHM vs. laser detuning
$\Delta$ from plot~(a), revealing the phenomenon of FWHM reduction
with increasing $\Delta$. Please note a reduced QD-FM detuning
$\Delta E = +230\,\mu$eV, yielding $T_1 \sim 410 \pm 20$~ps and
correspondingly larger FWHM by factor $\sim 1.6$ for $\Delta = 0$
conditions with respect to Fig.~2(d).\\


\begin{thebibliography}{99}
\bibitem{footnote} The center Mollow peaks have been omitted in our model calculations due to increasing masking by laser stray light.
\end{thebibliography}
\end{document}